# Andreev reflection of fractional quantum Hall quasiparticles

M. Hashisaka[1,2,*], T. Jonckheere[3], T. Akiho[1], S. Sasaki[1], J. Rech[3], T. Martin[3] & K. Muraki[1]

[1] NTT Basic Research Laboratories, NTT Corporation, 3-1 Morinosato-Wakamiya, Atsugi, Kanagawa 243-0198, Japan

[2]JST, PRESTO, 4-1-8 Honcho, Kawaguchi, Saitama 332-0012, Japan

[3]Aix Marseille Univ, Université de Toulon, CNRS, CPT, Marseille, France

*masayuki.hashisaka.wf@hco.ntt.co.jp

**Abstract**

Electron correlation in a quantum many-body state appears as peculiar scattering behaviour at its boundary, symbolic of which is Andreev reflection at a metal-superconductor interface. Despite being fundamental in nature, dictated by the charge conservation law, however, the process has had no analogues outside the realm of superconductivity so far. Here, we report the observation of an Andreev-like process originating from a topological quantum many-body effect instead of superconductivity. A narrow junction between fractional and integer quantum Hall states shows a two-terminal conductance exceeding that of the constituent fractional state. This remarkable behaviour, while theoretically predicted more than two decades ago but not detected to date, can be interpreted as Andreev reflection of fractionally charged quasiparticles. The observed fractional quantum Hall Andreev reflection provides a fundamental picture that captures microscopic charge dynamics at the boundaries of topological quantum many-body states.

When a two-dimensional electron system (2DES) is subjected to a perpendicular magnetic field at low temperatures, electrons condense into the strongly correlated phase of the fractional quantum Hall (FQH) state[1]. Quasiparticles in FQH systems have fascinating properties, such as fractional charge[2] and anyonic statistics[3]. Furthermore, for particular states such as that at Landau-level filling factor $v = 5/2$, theory predicts that quasiparticles obey non-Abelian braiding statistics that provide the basis of fault-tolerant quantum computation[4,5]. The fractional charge[6-9] and anyonic nature[10-12] of the quasiparticles have been revealed experimentally by shot-noise measurements and Fabry-Pérot interferometry. These studies have elucidated the behaviour of quasiparticles within the FQH state—either bulk or edges—that gives their defining properties. On the other hand, one may expect the quasiparticles to exhibit unique behaviour at an interface between the FQH state and another topologically distinct system, in a similar way as the Cooper-pair correlation in a superconductor manifests itself as Andreev reflection, where an electron incident from a normal metal to a superconductor is reflected as a hole[13,14]. This, in turn, poses a fundamental question as to whether electron correlation in a topological quantum many-body state shows up as a unique interface phenomenon. FQH Andreev reflection, which we demonstrate in this paper, is an elementary process that answers this question.

The FQH Andreev process has been predicted by theories examining charge transport across a narrow junction between quantum Hall (QH) states with different filling factors. The most intensively studied system is one comprised of the $v = 1/3$ Laughlin state and the $v = 1$ integer QH (IQH) state[15,16]. The charge transport can be modelled as the tunnelling between the $v = 1/3$ and 1 edge channels, which can be treated as a chiral Luttinger liquid and a Fermi liquid, respectively[17]. When the channels are coupled through a single scatterer, the problem can be solved analytically by transforming it into that of tunnelling between edge states with Luttinger parameter $g = 1/2$[18,19]. The exact solution predicts that in the strong-coupling regime the two-terminal conductance $G$ exceeds the conductance $e^2/3h$ ($e$: electron charge, $h$: Planck's constant) of the $v = 1/3$ state, reaching $e^2/2h$ in the strong-coupling limit[15,16,18-21]. The enhancement of $G$ can be interpreted as the result of the Andreev process, where two incoming charge-$e/3$ quasiparticles are scattered into a transmitted electron with charge $e$ and a reflected quasihole with charge $-e/3$[15]. This theoretical prediction, however, has not yet been confirmed experimentally, despite recent progress in experiments on related systems[22-26].

In this paper, we present evidence of the FQH Andreev process, namely $G$ exceeding $e^2/3h$ in a narrow junction between $v = 1/3$ and 1 states. As the junction width is varied using the split-gate voltage applied to form the junction, $G$ oscillates around $e^2/3h$, exhibiting several peaks where $G$ overshoots the bulk conductance $e^2/3h$, reaching $G \cong 1.2 \times e^2/3h$. The conductance oscillations indicate several Andreev processes at multiple scatterers present between the $v = 1/3$ and 1 edges. The evidence is also reinforced by demonstrating that the junction operates as a dc-voltage transformer generating a negative voltage output for positive input.

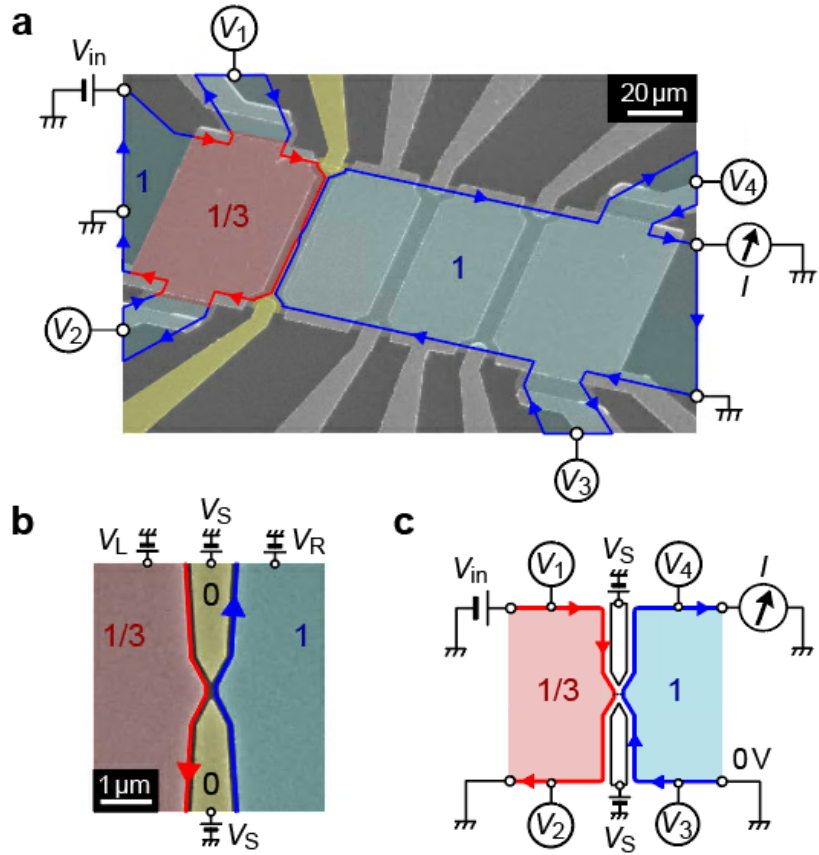

**Figure 1 Fractional-integer quantum Hall junction. a-b**, False-colour scanning electron micrograph of the Hall-bar sample with measurement configurations (**a**) and magnified view near the narrow junction (**b**). A perpendicular magnetic field $B = 9$ T is applied from the back to the front of the sample. The $v = 1$ states develop over the wide blue regions in the 2DES with the front-gate voltages, including $V_R$, set at 0 V. Meanwhile, electron density below one of the front gates (red region) is reduced to form the $v = 1/3$ state by applying $V_L = -0.42$ V. A narrow 1/3-1 junction is formed by depleting the 2DES under the split gate electrodes (yellow) with negative $V_S$. Chiral edge states are displayed by arrows (blue, between $v = 1$ and $v = 0$; red, between $v = 1/3$ and $v = 0$). $V_{in}$ is the applied source-drain voltage, $I$ is the measured current, and $V_i$ ($i = 1$-4) are the measured voltages of the incoming and outgoing channels of the 1/3-1 junction. **c**, Schematic of the experimental setup. A narrow junction is formed between $v = 1/3$ and $v = 1$ states.

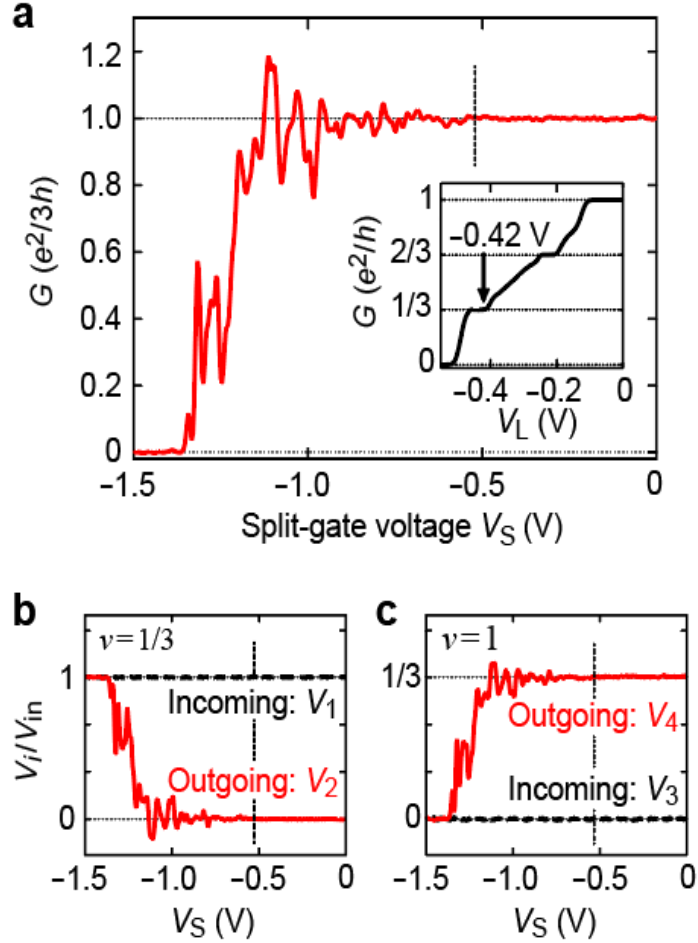

**Figure 2 Signatures of Andreev reflection. a**, Two-terminal conductance $G$ as a function of split-gate voltage $V_S$, taken with $V_L = -0.42$ V. Below $V_S \cong -0.55$ V (indicated by a dashed line), the 2DESs underneath the split gate electrodes are depleted to form a narrow 1/3-1 junction. Conductance oscillations with $G > e^2/3h$, the evidence of the Andreev reflection, are observed. (Inset) $G$ as a function of leftmost top gate voltage $V_L$ measured with $V_S = 0$, indicating that a $\nu = 1/3$ state forms at $V_L = -0.42$ V in the region immediately to the left of the junction (red region in Figs. 1a and 1b). **b-c**, $V_S$ dependence of the voltages on the incoming ($V_1$ and $V_3$) and outgoing ($V_2$ and $V_4$) channels, measured in the same setup as in **a** on the $\nu = 1/3$ (**b**) and $\nu = 1$ (**c**) sides of the junction. The vertical axes are normalised by the source-drain voltage $V_{in}$. Both negative output ($V_2 < 0$) in the reflected channel and overshoot ($V_4 > V_{in}/3$) in the transmitted channel are the signatures of the Andreev reflection.

Our QH device, formed in a Hall bar containing a 2DES in a GaAs quantum well, has several top gates and pairs of split gates in between (Fig. 1a). A perpendicular magnetic field of $B = 9$ T sets the bulk of the 2DES at $\nu = 1$. We then use the leftmost top gate ($V_L = -0.42$ V) to form a $\nu = 1/3$ region underneath (see the inset in Fig. 2a). A narrow 1/3-1 junction is formed by applying a negative gate bias $V_S$ to both electrodes of the split gate located immediately to the right of $v = 1/3$ region and depleting the 2DES underneath (Fig. 1b). In this situation, the setup for transport measurements can be expressed schematically as in Fig. 1c [see Supplementary Note 1]. We measured the two-terminal differential conductance $dI/dV_{in}$ by applying a source-drain voltage $V_{in} = V_{in}^{dc} + V_{in}^{ac}$ on the $v = 1/3$ side of the junction and measuring the transmitted current $I$ on the $v = 1$ side using a standard lock-in technique.

Figure 2a presents the central result of this paper, where we plot the zero-bias conductance $G$, i.e., $dI/dV_{in}$ at $V_{in}^{dc} = 0$ V, as a function of $V_S$. A narrow junction forms at $V_S < -0.55$ V. As $V_S$ is decreased below $-0.55$ V, the junction width decreases and $G$ starts to oscillate around $e^2/3h$ with the amplitude growing with decreasing $V_S$. The most striking observation is that $G$ overshoots $e^2/3h$ at several oscillation peaks before the junction is pinched off at $V_S \cong -1.4$ V. The maximum $G$ reaches $1.2 \times e^2/3h$ at $V_S \cong -1.1$ V. Such a two-terminal conductance, enhanced by narrowing the junction and exceeding the conductance of the constituent element, is nontrivial and counter-intuitive. We note that these features appear only in 1/3-1 junctions and not in 1/3-1/3 or 1-3 junctions (see Supplementary Note 7).

The peculiarity of the charge-transfer process is revealed alternatively by probing the potentials, or voltages $V_i$ ($i$ = 1-4) of the incoming and outgoing edge channels. Figures 2b and 2c display $V_i$ measured at $V_{in}^{dc} = 0$ V normalized by $V_{in}$, plotted as a function of $V_S$. The voltages $V_1$ and $V_3$ of the incoming channels are, respectively, equal to potentials $V_{in}$ and 0 V of the electrodes on their upstream, independent of $V_S$. In contrast, the voltages $V_2$ and $V_4$ of the outgoing channels vary with $V_S$. The most remarkable feature is the negative voltage that appears in $V_2$. Phenomenologically, this demonstrates that the junction operates as a dc-voltage transformer generating negative voltage output ($V_2 < 0$) for a positive input ($V_{in} > 0$).

From the Landauer-Büttiker formalism, $V_2$ and $V_4$ are related to $G$ as

$$V_2 = [1 - G(e^2/3h)^{-1}]V_{in}, \quad (1)$$

$$V_4 = G(e^2/3h)^{-1} V_{in}. \quad (2)$$

These formulas show that both $V_2 < 0$ and $V_4 > V_{in}/3$ correspond to $G > e^2/3h$. Within the picture of Andreev reflection, the negative voltage ($V_2 < 0$) of the back-reflected channel is a direct manifestation of the quasihole reflection.

While it is evident that the Andreev reflection is responsible for the observed $G > e^2/3h$, to understand microscopic processes therein, we need to explain the origin of the conductance oscillations, which is not predicted from the original models based on the tunnelling through a single scatterer[18,19]. Resonant tunnelling through unintentional discrete levels in the junction, which is responsible for the oscillations in the low-conductance regime near $V_S$ = −1.3 V (see Supplementary Note 4), cannot account for the conductance oscillations with $G$ > $e^2/3h$. In the following, we present the dependence of the conductance on $V_{in}$ and temperature $T$ and discuss the oscillation mechanism.

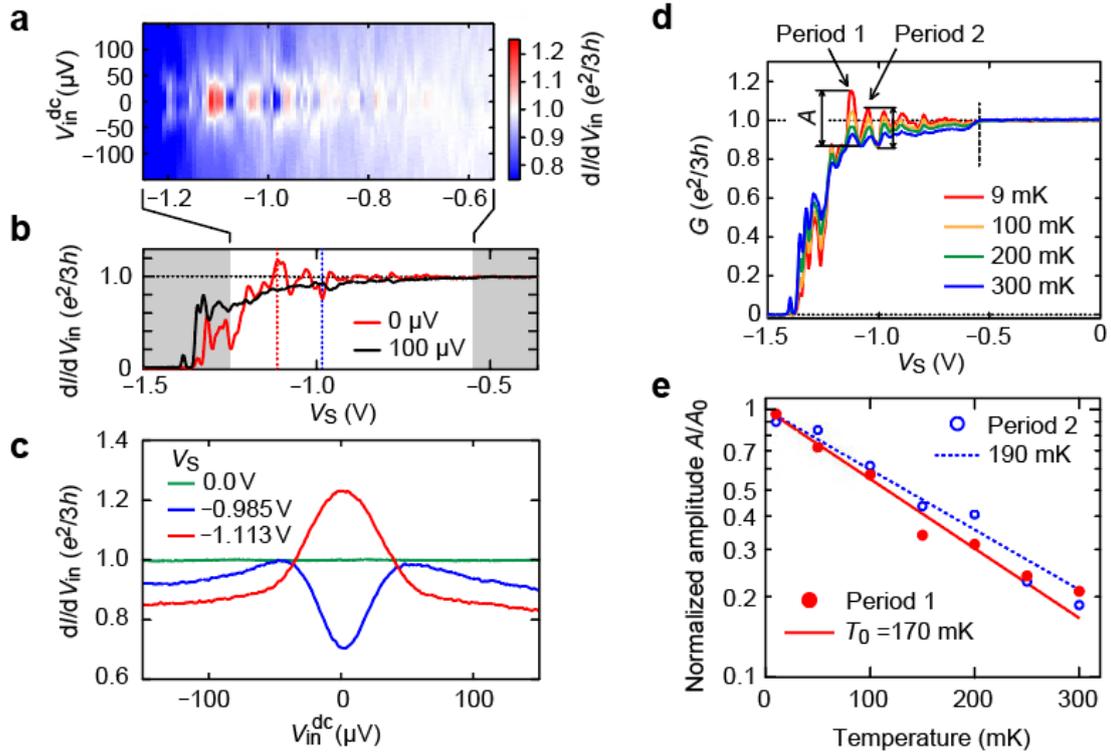

**Figure 3 Source-drain-voltage and temperature dependence of conductance oscillations. a**, Colour plot of d$I$/d$V_{in}$ as a function of $V_S$ and $V_{in}^{dc}$, indicating oscillations with d$I$/d$V_{in}$ > $e^2/3h$ at $|V_{in}|$ < 40 μV. **b**, Comparison of d$I$/d$V_{in}$ vs $V_S$ traces at zero and finite bias, indicating that the feature d$I$/d$V_{in}$ > $e^2/3h$ is absent at $V_{in}^{dc}$ = 100 μV. **c**, $V_{in}^{dc}$ dependence of d$I$/d$V_{in}$ at several $V_S$. Zero-bias conductance enhancement (suppression) is observed at $V_S$ = −1.113 V (−0.985 V), whereas no feature is seen without forming a narrow junction ($V_S$ = 0 V). **d**, $G$ vs. $V_S$ traces measured at several temperatures. The amplitudes $A$ of the oscillations in periods 1 and 2 are estimated as the peak-to-valley values. **e**, Temperature dependence of $A/A_0$ for periods 1 and 2. The curves of the form exp(−$T/T_0$) well fit the data using similar characteristic temperatures ($T_0$ = 170 and 190 mK).

Figure 3a displays a colour plot of differential conductance d$I$/d$V_{in}$ as a function of $V_S$ and $V_{in}^{dc}$. The oscillations with d$I$/d$V_{in}$ > $e^2/3h$ are seen only at $|V_{in}|$ < 40 μV. For illustration, we plot in Fig. 3b the pinch-off trace at $V_{in}^{dc}$ = 100 μV, where d$I$/d$V_{in}$ < $e^2/3h$ over the entire range of $V_S$. Figure 3c shows the $V_{in}^{dc}$ dependence of d$I$/d$V_{in}$ at $V_S$ = −1.113 and −0.985 V. At $V_S$ = −1.113 (−0.985) V, which corresponds to the peak (valley) of the oscillations in Fig. 3b, we observe a pronounced zero-bias enhancement (suppression) of the conductance. In contrast, at $V_S$ = 0 V, where the $v$ = 1/3 and 1 regions form a long junction spanning across the 80-μm-wide Hall bar, d$I$/d$V_{in}$ remains constant at $e^2/3h$. These results clearly show that the Andreev process is observed only in narrow junctions at low bias. The data also reveal that not only the conductance enhancement but also its suppression are low-bias anomalies.

Figure 3d shows the $T$ dependence of the conductance oscillations. The oscillation amplitude decreases with increasing $T$, and the signature of the Andreev process, $G > e^2/3h$, disappears above 200 mK. We focus on two single periods of the oscillations near $V_S$ = −1.113 and −1.095 V and extract the amplitude $A$ as the peak-to-valley value of $G$ in each period. The two sets of $A$ vs. $T$ data are well fitted by an exponential function $A_0\exp(-T/T_0)$, as shown in Fig. 3e, where $A_0$ is the amplitude at $T$ = 0 and $T_0$ is the characteristic temperature. The exponential temperature dependence bears analogy with that seen in various electronic interferometers[27]. The $T_0$ values (170 and 190 mK) are close to each other, indicating that these oscillations share the same origin in nature. The data in Fig. 3 also demonstrate that the conductance oscillations with $G > e^2/3h$ are highly reproducible (similar conductance oscillations were reproduced for different cool-downs and in different samples, see Supplementary Note 3, 5, and 6).

We argue that several Andreev processes in the junction are responsible for the conductance oscillations, as predicted in theories involving multiple scatterers or a line junction of finite width[18,19,28-32]. We consider $N$ scatterers along the counter-propagating $v$ = 1/3 and 1 channels and incoherent transport between them. $N$ is proportional to the junction width (i.e., length of the counter-propagating channels) and hence varies with $V_S$. Here, "incoherent transport" means that the $N$ scatterers give independent scattering events, where the outgoing channels are characterized by a chemical potential that defines the input for the next scatterer. With this assumption the voltage of the $v$ = 1/3 (1) channel incoming to the $n$th scatterer is given by $V_{n-1} = i_{n-1} \times 3h/e^2$ ($W_{N-n} = j_{N-n} \times h/e^2$), where $i_{n-1}$ ($j_{N-n}$) is the current in the incoming channel. With this setup, one can use the exact solution for the single-scatterer model[19] to define the conductance $g_n$ for each scatterer as a function of the applied bias $V_{n-1} - W_{N-n}$ and evaluate the current $I_n$ flowing through it. Notably, as shown for the single-scatterer case, the charge conservation law and the requirement that the outgoing power be equal to or less than the incoming one lead to $0 \leq g_n \leq 1/2$[21]. Here, the charge transport through the $n$th scatterer becomes dissipationless only when $g_n$ = 0 or 1/2. The latter (former) corresponds to the strong-coupling limit (complete decoupling). Namely, tunnelling for any intermediate $g_n$ values is accompanied by energy dissipation.

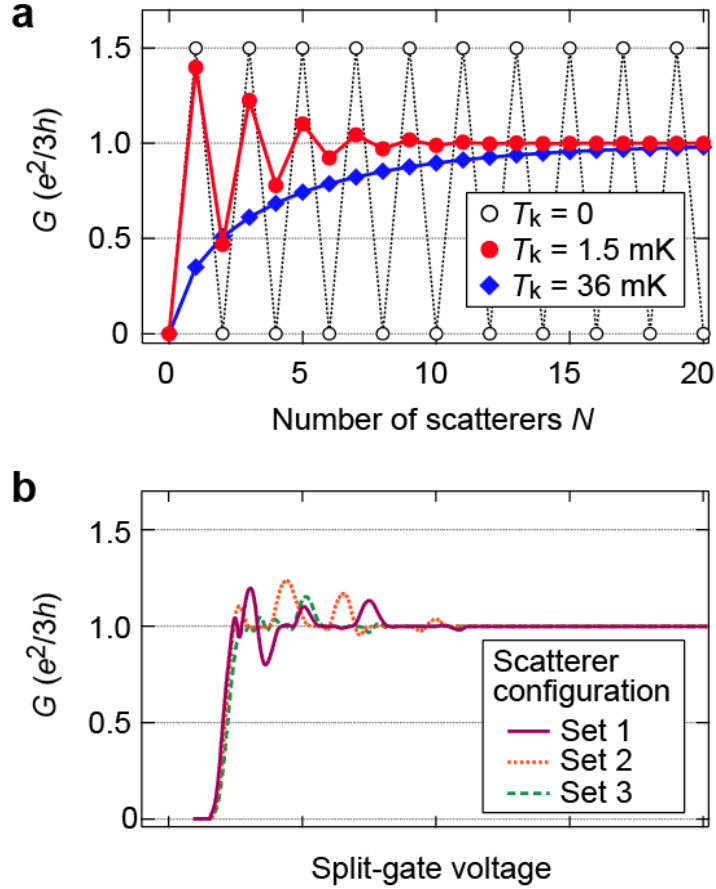

**Figure 4 Simulation of conductance oscillations using incoherent $N$ scatterers model. a**, $N$ dependence of $G$ for several coupling strengths: strong-coupling limit ($T_k = 0$, black open circles), intermediate ($T_k = 1.5$ mK, red filled circles), and weak couplings ($T_k = 36$ mK, blue diamonds). **b**, Simulations taking the effects of the confining potential and the randomness in the positions of the scatterers into account. A window function was used to relate the split-gate voltage with the number and strength of scatterers (for details, see Supplementary Note 8). The three traces (1-3) are obtained for different configurations of the scatterers' positions.

The conductance $G$ of the whole junction is obtained by solving a non-linear system of equations numerically. The results are shown in Fig. 4a for three representative cases: strong ($T_k = 0$, black open circles), intermediate ($T_k = 1.5$ mK, red filled circles), and weak couplings ($T_k = 36$ mK, blue diamonds) under the experimental condition with an applied voltage of 20 μV and a temperature of 9 mK. Here, $T_k$ is the crossover energy scale between strong- and weak-coupling regimes[19] (for details, see Supplementary Note 8). In the strong-coupling limit, where we have $g_n = 1/2$ for all $n$, each scatterer only switches the sign of the voltage between the channels without causing energy dissipation.

Consequently, $G$ oscillates as a function of $N$ between 0 ($N$ even) and $e^2/2h$ ($N$ odd) (black circles). This oscillation can be regarded as the result of successive dissipationless Andreev processes, where the tunnel current switches direction at each scatterer without changing magnitude. When the coupling weakens to give $g_n < 1/2$, each scatterer equilibrates the channels, which results in the reduced output voltage at each scatterer and hence damping of the conductance oscillations (red circles). The damping is significant particularly for large $N$, where the channels experience equilibration many times. When the coupling weakens further to give $g_n < 1/3$ for all $n$, the tunnel current flows only in one direction. In this case, $G$ monotonically increases with $N$, asymptotically approaching $G = e^2/3h$[30] (blue diamonds). The simulation for the intermediate coupling (red circles) captures essential features of the experimental data in Fig. 2a. If we take into account more realistic experimental situations, including the confining potential of the split gate and randomness in the positions of the scatterers, the simulation can even better reproduce the experimental features (Fig. 4b). In the simulation, the confining potential controls the effective width of the junction by multiplying a position-dependent window function to the coupling strength of the scatterers, resulting in weaker coupling near the junction ends and hence reduced oscillation amplitude (for details, see Supplementary Note 8).

While the above multiple-scatterer model well explains the $V_S$ dependence of $G$, it still fails to account for the observed $V_{in}$ and $T$ dependence. Since the model inherits the $V_{in}$ and $T$ dependence of the conductance from the single-scatterer model[18,19], which gives $dI/dV_{in}$ as an increasing function of $V_{in}$ and $T$, it remains incapable of reproducing oscillations decaying with $V_{in}$ or $T$. This, in turn, suggests that coherent processes neglected in the above model, such as interference between successive scattering events[33], play an important role. We speculate that constructive (destructive) interference of tunnelling amplitudes can enhance (suppress) the coupling strengths of several scatterers in some range of $V_S$. Indeed, a theory considering coherent interference predicts that a 1/3-1 junction with Coulomb interaction shows conductance oscillations up to $e^2/2h$ as a function of the junction width[28]. In this view, conductance enhancement or suppression at the extrema of the oscillations is partly due to the interference enhancement of the coupling. This picture explains why the oscillations appear only at low $V_{in}$ and $T$. Furthermore, it helps to understand why the simulation for the weak-coupling regime of the incoherent model can mimic the $V_S$ dependence of $G$ at high $V_{in}$ (Fig. 3b) or high $T$ (Fig. 3d).

Finally, we discuss a related interesting issue, namely the mixing of the $v = 1/3$ and 1 edge modes expected for a wide junction. Counter-propagating $v = 1/3$ and 1 channels studied here is a basic setup in the model for the edge modes of the hole-conjugate $v = 2/3$ FQH state[34]. There, inter-channel Coulomb interaction and disorder-assisted tunnelling govern the mixing of the channels and thus determine their fate in the low-temperature and long-channel limit. The conductance oscillations with $G$ exceeding $e^2/3h$ observed in our experiment can be interpreted as a precursor phenomenon of the mixing process, namely the "mesoscopic fluctuation" predicted in Ref. 32, suggesting the presence of the neutral-mode physics of counter-propagating channels at the 1/3-1 junction[31,32]. Our findings indicate that the Andreev process is a vital ingredient therein.

We have demonstrated FQH Andreev reflection, which is one of the essential concepts for understanding edge transport at the boundaries of topological quantum many-body systems. We expect to observe similar Andreev processes in various FQH junctions with different electronic systems, including non-quantum Hall systems such as normal metals and superconductors[35,36].

## Methods

**Sample fabrication.** We fabricated the sample in a 2DES in a GaAs quantum well of 30 nm width. The centre of the well is located 190 nm below the surface. The sample was patterned using e-beam lithography for fine gate structures and photolithography for chemical etching, coarse metalized structures, and ohmic contacts formed by alloying Au-Ge-Ni on the surface.

**Measurement setup.** We set electron density in the 2DES at $2.2 \times 10^{11}$ cm$^{-2}$ by applying a back-gate voltage of 1.29 V at a refrigerator temperature of 9 mK, except for the data in Figs. 3d and 3e. A perpendicular magnetic field $B = 9$ T was applied from back to front of the sample. The lock-in measurements were performed with the ac modulation of $V_{in}^{ac} = 20$ μV RMS at 31 Hz. The experimental results demonstrated in the main text were obtained for the split-gate device with the opening of 300 nm. The data from the devices with wider apertures are available in Supplementary Note 6.

**Acknowledgements**

The authors thank T. Ito, N. Shibata, and T. Fujisawa for fruitful discussions and H. Murofushi for technical support. This work was supported by Grants-in-Aid for Scientific Research (Grants No. JP16H06009, No. JP15H05854, No. JP26247051, and No. JP19H05603) and JST PRESTO Grant No. JP17940407.

**Author contributions** M.H. conceived the experiment. M.H., T.A., and S.S. fabricated the sample. M.H. performed the measurement and analysed the data. T.J. and K.M. performed the simulations. M.H., T.J., and K.M. interpreted the results with help from J.R. and T.M. M.H. wrote the manuscript with help from T.J. and K.M. All authors discussed the results and commented on the manuscript.

# Supplementary Information for
# "Andreev reflection of fractional quantum Hall quasiparticles"

M. Hashisaka[1,2], T. Jonckheere[3], T. Akiho[1], S. Sasaki[1], J. Rech[3], T. Martin[3] and K. Muraki[1]

[1]NTT Basic Research Laboratories, NTT Corporation, 3-1 Morinosato-Wakamiya, Atsugi, Kanagawa 243-0198, Japan
[2]JST, PRESTO, 4-1-8 Honcho, Kawaguchi, Saitama 332-0012, Japan
[3]Aix Marseille Univ, Université de Toulon, CNRS, CPT, Marseille, France

**Supplementary Note 1: Landauer-Büttiker edge transport picture**

Supplementary Fig. 1 is a schematic of the experimental setup showing the whole of the Hall-bar device, where the blue and red arrows show the $v = 1$ and $1/3$ edge channels, respectively. When the counter-propagating $v = 1$ and $1/3$ channels are fully equilibrated at the wide junction across the Hall bar, a chiral one-dimensional channel of conductance $2e^2/3h$ is formed at the junction, as shown by a black arrow. The channels incoming to the narrow junction do not experience equilibration with any other channels before impinging on it, because both bulk $v = 1$ and $1/3$ states are insulating (incompressible). Therefore, the incoming voltages $V_1$ and $V_3$ correspond to the voltages of the electrodes at their upstream. Actually, we observe $V_1 = V_{in}$ and $V_3 = 0$ over the entire range of $V_S$, as shown in Figs. 2b and 2c. When the narrow junction has conductance $g$, the transmitted current $I$ is given by $I = g(V_1 - V_3) = gV_{in}$; hence, we find $g = G$. This justifies the picture in Fig. 1c. Resultantly, the outgoing voltages $V_2$ and $V_4$ are expressed as $V_2 = V_{in} - I \times (3h/e^2) = [1 - G(e^2/3h)^{-1}]V_{in}$ and $V_4 = I \times (h/e^2) = G(e^2/h)^{-1}V_{in}$, as described in the main text.

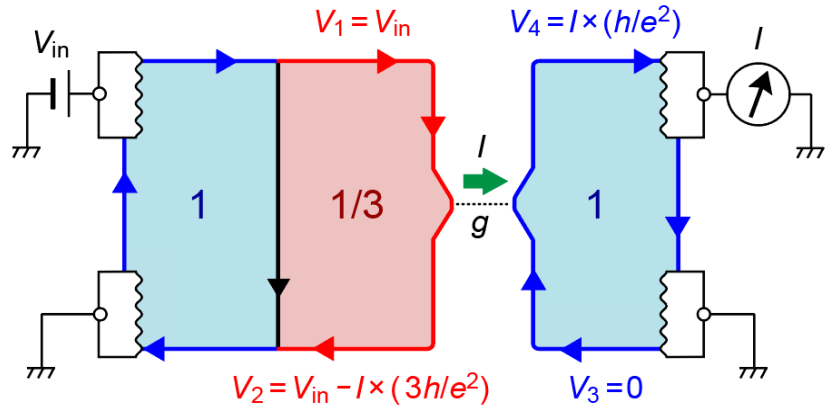

**Supplementary Figure 1.** Schematic of experimental setup.

# Supplementary Note 2: Longitudinal and vertical resistances of the two-dimensional electron system

Supplementary Fig. 2a shows the longitudinal ($R_{xx}$) and vertical ($R_{xy}$) resistances of the bulk two-dimensional electron system (2DES) as a function of the perpendicular magnetic field $B$ at the back-gate voltage $V_{BG} = 1.29$ V. Supplementary Fig. 2b shows a colour plot of $R_{xx}$ in the $V_{BG}$-$B$ plane. The experimental results presented in the main text were obtained at $B = 9.0$ T and $V_{BG} = 1.29$ V ($\nu \cong 1$), indicated by the dotted line in Supplementary Fig. 2a and the white circle in Supplementary Fig. 2b.

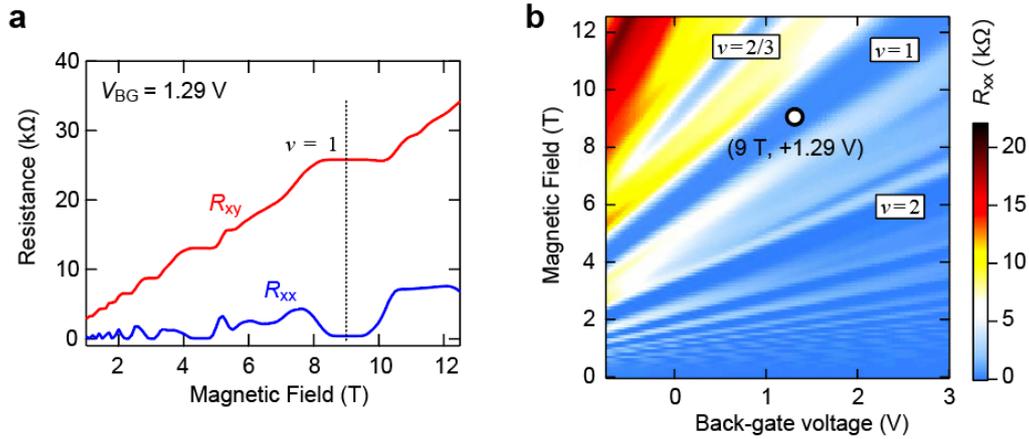

**Supplementary Figure 2. a,** Magnetic field dependence of $R_{xx}$ and $R_{xy}$. **b,** Colour plot of $R_{xx}$ as a function of $V_{BG}$ and $B$.

# Supplementary Note 3: Asymmetric split-gate biasing

Supplementary Figs. 3a and 3b show the colour plots of $G$ measured as a function of the gate voltages $V_{S1}$ and $V_{S2}$ independently applied to the upper and lower split-gate electrodes (see Supplementary Fig. 3d). The two graphs show the same experimental results in different ranges: $0.9 \times e^2/3h \leq G \leq 1.1 \times e^2/3h$ in Supplementary Fig. 3a; $0 \leq G \leq 1.25 \times e^2/3h$ in Supplementary Fig. 3b. Whereas $G = e^2/3h$ at $V_{S1} > -0.55$ V and/or $V_{S2} > -0.55$ V, $G$ deviates from $e^2/3h$ when both $V_{S1}$ and $V_{S2}$ are below $-0.55$ V, because the 2DESs under the split-gate metals are depleted near $-0.55$ V. In this area, we observe the conductance oscillations with $G > e^2/3h$ (shown in red in Supplementary Fig. 3a) that are the signatures of several Andreev processes. Some peaks and dips in the conductance oscillations are likely to appear parallel to either $V_{S1}$ or $V_{S2}$ axis [e.g. along the line (i) in Supplementary Fig. 3b], suggesting that the number of scatterers in the junction decreases one by one with these gate voltages. Supplementary Fig. 3c shows $G$ traces as a function of the split-gate voltages swept along the four lines in Supplementary Fig. 3b. We observe a variety of conductance oscillations that reflect the positions of scatterers in the 1/3-1 junction.

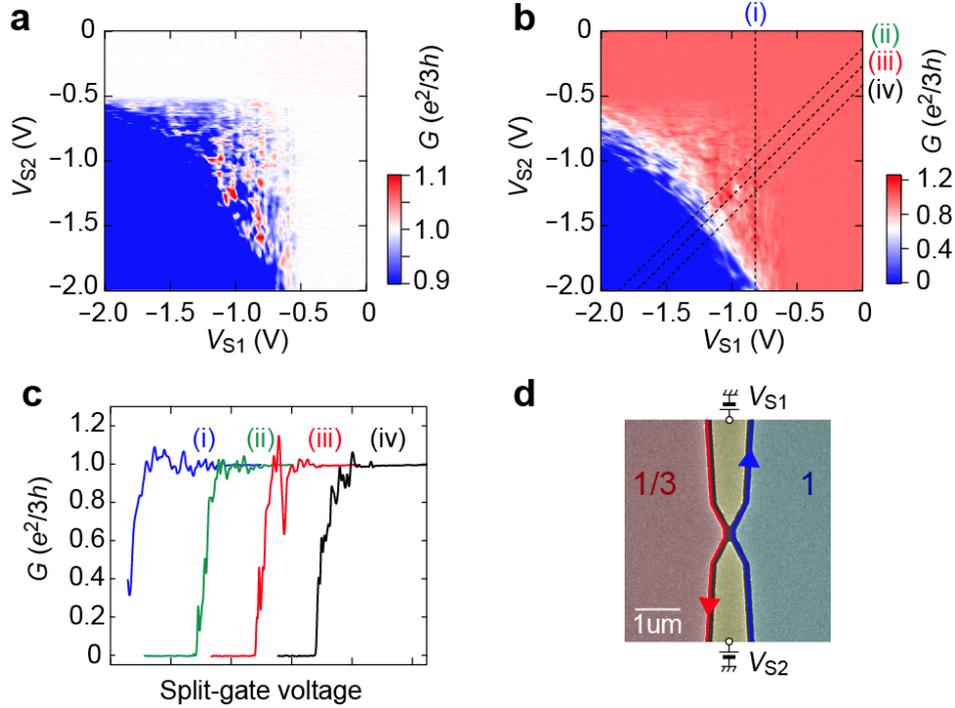

**Supplementary Figure 3. a,b,** Colour plots of $G$ as a function of $V_{S1}$ and $V_{S2}$ in different $G$ ranges: $0.8 \times e^2/3h \leq G \leq 1.2 \times e^2/3h$ in **a** and $0 \leq G \leq 1.25 \times e^2/3h$ in **b**. **c,** Pinch-off characteristics of $G$ obtained by the split-gate voltage sweeps along the dashed lines in **b** (horizontally shifted for clarity). **d,** Gate voltages $V_{S1}$ and $V_{S2}$ applied to the split gate.

## Supplementary Note 4: Resonant tunnelling through discrete levels

While our experimental results can be well understood with the multiple-scatterer model, here we complement our discussion by excluding the possibility that the conductance oscillations originate from resonant tunnelling through unintentionally formed discrete levels. Such discrete levels could form when puddles of different filling factors exist near the junction[1]. First of all, the presence of a discrete level by itself cannot induce $G > e^2/3h$, because in the absence of the Andreev process the conductance is upper-limited to $e^2/3h$ of the $\nu = 1/3$ region. Second, the increase in the oscillation amplitude with decreasing $V_S$ (Fig. 2a) is inconsistent with the $V_S$ dependence of the tunnel barrier height that increases with decreasing $V_S$. Third, when we independently sweep the split-gate voltages $V_{S1}$ and $V_{S2}$ (Supplementary Fig. 3d), no features of resonant tunnelling are seen in the region of $G > 0.8 \times e^2/3h$ (Supplementary Fig. 3a); the conductance peaks and dips are likely to appear parallel to either $V_{S1}$ or $V_{S2}$. This observation indicates that the position of the scatterers within the junction, not the energy level, is essential. This contrasts with the conductance peaks near the pinch-off where $G$ is well below $e^2/3h$; they shift diagonally with the asymmetric biasing, which is the behaviour expected for resonant levels (see Supplementary Fig. 3b). Thus, we conclude that the conductance oscillations around $G = e^2/3h$ are not caused by the resonant tunnelling but by the Andreev processes in the multiple-scatterer system.

## Supplementary Note 5: Magnetic field dependence

We examined the $B$ dependence of the pinch-off characteristics of the 1/3-1 junction. Supplementary Fig. 4a shows a colour plot of $G$ in the $V_S$-$B$ plane. The conductance oscillations with $G > e^2/3h$ are observed over the measured range of 8.8 T $< B <$ 9.2 T, indicating the Andreev processes remain present with a slight change in $B$. Near $V_S = -1.113$ V, $G$ seems to oscillate as a function of $B$ showing $G > e^2/3h$ at several peaks (Supplementary Fig. 4b). The oscillating $B$ dependence may be related to the Aharonov-Bohm interference between tunnelling amplitudes through different scatterers.

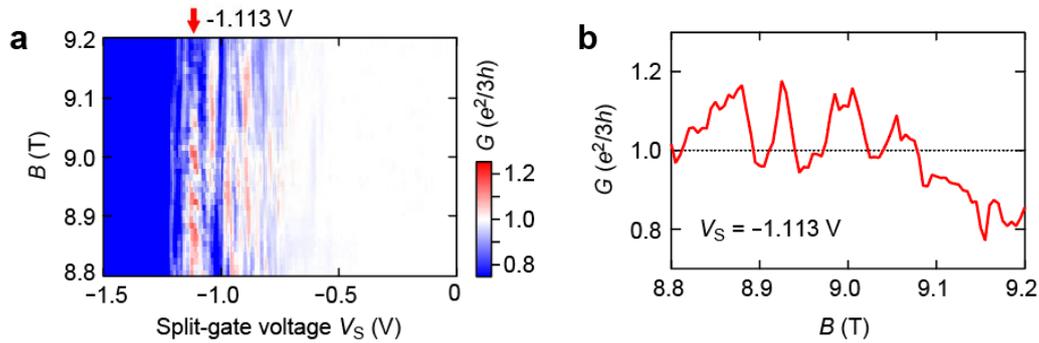

**Supplementary Figure 4. a,** Colour plot of $G$ as a function of $V_S$ and $B$. **b,** $B$ dependence of $G$ at $V_S = -1.113$ V.

## Supplementary Note 6: Different samples

The signature of the Andreev process, $G > e^2/3h$, is observed not only in the split-gate device demonstrated in the main text, which has the split-gate opening of 300 nm, but also in other samples having wider openings. We examined two samples, one with 600 and the other with 900-nm openings, fabricated on the same wafer. Supplementary Fig. 5 shows the pinch-off characteristics of these samples. Like the 300-nm device, both of them show the conductance oscillations with $G > e^2/3h$, manifesting the Andreev reflection. It is worth noting that their oscillation amplitudes are smaller than those for the 300-nm device, suggesting enhanced equilibration or weaker couplings due to strong negative $V_S$ in the wider samples.

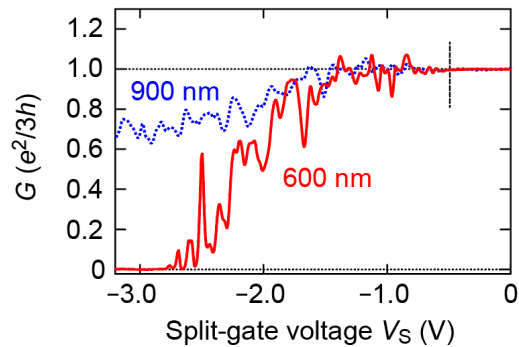

**Supplementary Figure 5.** Signatures of the Andreev reflection in different samples.

## Supplementary Note 7: Transport properties of different quantum Hall junctions

It is instructive to look at the transport properties of other QH junctions having different sets of filling factors. Supplementary Fig. 6a shows the pinch-off characteristics of a junction between $v = 1/3$ states formed by applying the top-gate voltages $V_L = V_R = -0.42$ V at $V_{BG} = 1.29$ V and $B = 9$ T. Note that $G$ never exceeds the conductance $e^2/3h$ over the entire range of $V_S$. At $V_S > -0.55$ V, where the 2DES under the split gate is not depleted, we observe $G < e^2/3h$, which results from the enhanced backscattering through puddles of different filling factors. Below $V_S = -0.55$ V, the conductance through the narrow junction is maintained at $G \cong e^2/3h$ down to $V_S = -0.95$ V. Below $-0.95$ V, it decreases to zero, indicating that the junction is completely pinched off at $V_S \cong -1.2$ V. Supplementary Fig. 6b displays another result obtained from an IQH junction between $v = 1$ and $v = 3$ states. The system was prepared by applying $V_L = -0.42$ V at $V_{BG} = 1.29$ V and $B = 3$ T. Likewise, in this case, $G$ decreases from $e^2/h$ to zero without showing $G > e^2/h$ over the entire range of $V_S$. Thus, the signature of Andreev reflection is neither observed for a junction between the same $v = 1/3$ states nor between different IQH states. This, in turn, clearly exhibits that the conductance oscillations with $G > e^2/3h$, demonstrated in the main text, are responsible for the Andreev reflection at the 1/3-1 junction.

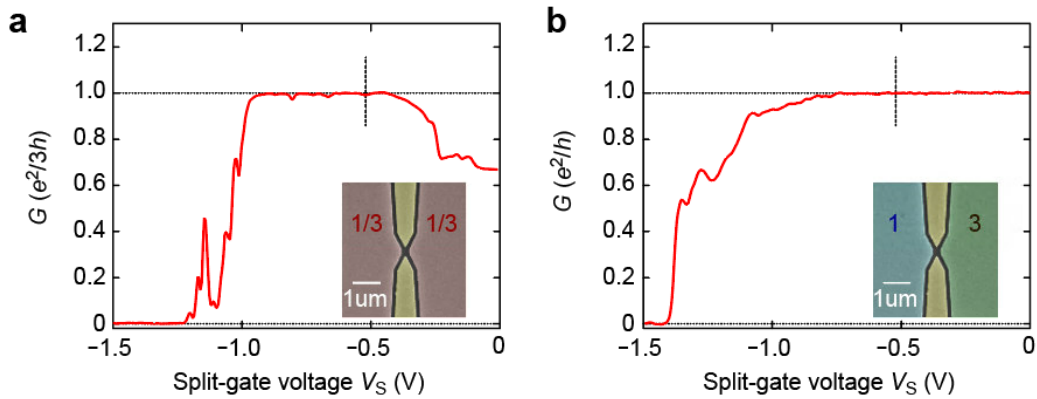

**Supplementary Figure 6. a,** $V_S$ dependence of $G$ of junction between $v = 1/3$ states. **b,** $V_S$ dependence of $G$ of junction between $v = 1$ and $v = 3$ states.

**Supplementary Note 8: Theoretical discussion**

The tunnelling problem through a single scatterer between $v = 1/3$ and $v = 1$ edge states has been studied and solved in the literature[2]. By using a Luttinger liquid representation for the edge states, one can show that this system can be mapped on a junction between two $v = 1/2$ edge states, which can be solved exactly using a refermionisation procedure. The total current through the junction, for a bias voltage $V_{in}$ and temperature $T$, can be written as

$$I(V_{in}, T, T_k) = \frac{e^2}{2h} V_{in} - \frac{e}{4h} T_k \sinh\left(\frac{V_{in}}{2T}\right) F_3\left(\frac{T_k}{2T}, \frac{eV_{in}}{2T}\right), \quad (1)$$

where $T_k$ parametrises the coupling strength of the scatterer, and

$$F_3(a,b) = \frac{\pi}{\cosh(ia+b)\cosh(ia-b)} - \frac{i}{\sinh(2b)}\left[\psi\left(\frac{1}{2} - \frac{a+ib}{\pi}\right) + \psi\left(\frac{1}{2} + \frac{a+ib}{\pi}\right) - \psi\left(\frac{1}{2} - \frac{a-ib}{\pi}\right) - \psi\left(\frac{1}{2} + \frac{a-ib}{\pi}\right)\right], \quad (2)$$

with $\psi$ the digamma function.

The most important result is that the conductance of the junction is $e^2/2h$ in the strong-coupling limit (corresponding to $T_k \to 0$), and the effective quasiparticles in the transport process are collective excitations with a charge $e/2$, which is different from the charge of the individual excitations existing at each edge. The fact that the conductance reaches $e^2/2h$ means that the output voltage on the $v = 1/3$ can be larger than the input one and can be understood in terms of Andreev reflection.

When the junction between the two edge states is wide, it cannot be modelled as a single scatterer, and a model with many scatterers can be used. While the problem becomes much more complicated, one regime where results can be obtained easily is the incoherent regime (we defined "incoherent" in the main text). There, all interference effects are neglected, and the current through a given scatterer depends only on the incoming voltage on the two edges. The incoherent multiple-scatterer model is illustrated in Supplementary Fig. 7 for the case of four scatterers. The voltages $V_1, \ldots V_N$ and $W_1, \ldots W_N$ can be obtained by solving the non-linear system of equations for $n = 1, \ldots N$:

$$V_n = V_{n-1} - \frac{3h}{e^2} I_n(V_{n-1} - W_{N-n}, T, T_{k,n}), \quad (3)$$

$$W_n = W_{n-1} + \frac{h}{e^2} I_{N-n+1}(V_{N-n} - W_{n-1}, T, T_{k,N-n+1}), \quad (4)$$

where $I_n$ is given by Eq. (1), and $V_0$ and $W_0$ are the incoming voltages. One can solve the Eqs. (3) and (4) numerically to obtain the outgoing voltages $V_N$ and $W_N$, or the output currents from the junction.

When the scatterers are weak, the system can be linearized and solved exactly by going to the continuum limit[3]. For the conductance as a function of the junction width, or the length $L$ of the counter-propagating channels, one obtains

$$G(L, l) = \frac{e^2}{3h} \times \frac{1 - \exp(-2L/l)}{1 - (1/3)\exp(-2L/l)}, \quad (5)$$

where $l$ is an effective equilibration length, which is related to the coupling strength. This result shows that, in the regime of many weak scatterers, the conductance never goes above $e^2/3h$, and it reaches this value on a length scale $\sim l$. This is the behaviour shown in Fig. 4a with the blue diamonds obtained from the numerical calculation ($T_k = 36$ mK). It is also the behaviour observed experimentally, for the conductance as a function of the split-gate voltage at high $V_{in}$ or $T$ (where oscillations are washed out), in Figs. 3b and 3d. A very different regime is obtained in the strong-coupling limit, where $T_{k,n}$ go to zero to give $g_n = 1/2$ for all $n$. There, $G$ becomes $e^2/2h$ for an odd $N$ and 0 for an even $N$, as shown by the black open circles in Fig. 4a. Between these regimes, for scatterers strong enough, a typical behaviour is shown by the red filled circles in Fig. 4a ($T_k = 1.5$ mK), namely, as a function of $N$, the conductance $G$ oscillates around the value $e^2/3h$, with $G > e^2/3h$ for odd $N$, and $G < e^2/3h$ for even $N$. The amplitude of these oscillations decreases as $N$ increases, and $G$ reaches eventually $e^2/3h$ for large $N$.

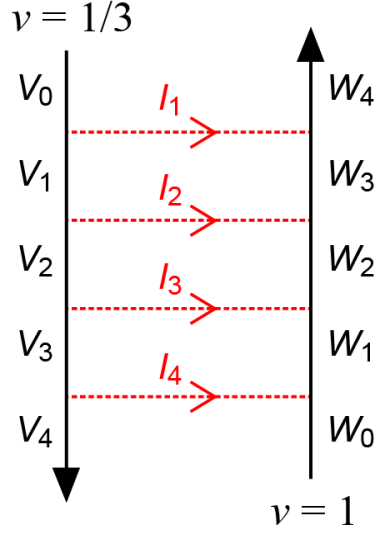

**Supplementary Figure 7.** Multiple-scatterer model, shown for the case of $N = 4$ scatterers.

In the experiment, one does not have direct access to the number of scatterers. However, the number of scatterers should be roughly proportional to the junction width, which is controlled by the applied split-gate voltage $V_S$. We model the continuous $V_S$ dependence of the conductance by making the following reasonable assumptions:

- When no split-gate voltage is applied, the junction consists of a large number $N$, randomly placed inside the junction. In practice, we put 40 scatterers, with equal strength $T_k$, inside the range $[-10, 10]$ on the $x$-axis.

- The effect of $V_S$ is modelled as a window function which reduces the effective width of the junction, with the strength of each scatterer going smoothly from $T_k$ to $\infty$ (= zero-strength scatterer) when the scatterer position goes from inside to outside the effective width determined by $V_S$. In practice, the strength of a given scatterer at position $x$ is provided by the following function of $V_S$:

$$T_k(V_s) = \frac{T_k}{[1-f(x-V_s)]f(x-V_s)}, \qquad (6)$$

where $f$ is a sigmoid function

$$f(y) = \frac{1}{1+\exp(-y/w)}, \qquad (7)$$

where $w$ defines the width of the transition region of the window function (in practice, we have chosen $w = 0.2$). With this choice of window function, all scatterers are suppressed for negative $V_S$ (corresponding to completely pinched-off junction), and the number of active scatterers increases as $V_S$ increases, with all scatterers fully active when $V_S > 10$.

Choosing a relatively strong coupling strength $T_k = 1.1$ mK, with an applied voltage 20 µV at temperature 9 mK (for a single scatterer, these parameters give the conductance ~ 0.49 $e^2/h$), we obtain the curves in Fig. 4b, each curve corresponding to a different random realization of the position of the scatterers. One can see that the major qualitative features of the experimental results are present; the amplitude of conductance oscillations around $e^2/3h$ decreases as $V_S$ is increased, while the inherent oscillating pattern depends on the fine details of the positions of the scatterers.